\documentclass[aip,twocolumn,floatfix,reprint]{revtex4-1}

\usepackage[T1]{fontenc} 
\usepackage{amsmath}
\usepackage{amsfonts}
\usepackage{amssymb}
\usepackage{graphicx}
\usepackage{textcomp}
\usepackage{mathtools}

\newcommand{\mpc}{\,\mathrm{mol}\,\%}

\begin{document}

\title{Soft particles at liquid interfaces: From molecular particle architecture to collective phase behavior}

\author{Simone Ciarella}
\affiliation{Department of Applied Physics, Eindhoven University of Technology, 5600 MB Eindhoven, The Netherlands}
\author{Marcel Rey}
\affiliation{Institute of Particle Technology, Friedrich-Alexander University Erlangen-N\"urnberg, Cauerstrasse 4, 91058 Erlangen, Germany}
\author{Johannes Harrer}
\affiliation{Institute of Particle Technology, Friedrich-Alexander University Erlangen-N\"urnberg, Cauerstrasse 4, 91058 Erlangen, Germany}
\author{Nicolas Holstein}
\affiliation{Institute of Particle Technology, Friedrich-Alexander University Erlangen-N\"urnberg, Cauerstrasse 4, 91058 Erlangen, Germany}
\author{Maret Ickler}
\affiliation{Institute of Particle Technology, Friedrich-Alexander University Erlangen-N\"urnberg, Cauerstrasse 4, 91058 Erlangen, Germany}
\author{Hartmut L\"owen}
\affiliation{Institute for Theoretical Physics II: Soft Matter, Heinrich-Heine University D\"usseldorf, D-40225 D\"usseldorf, Germany}
\author{Nicolas Vogel}
\affiliation{Institute of Particle Technology, Friedrich-Alexander University Erlangen-N\"urnberg, Cauerstrasse 4, 91058 Erlangen, Germany}
\email{nicolas.vogel@fau.de}
\author{Liesbeth M.~C.~Janssen}
\affiliation{Department of Applied Physics, Eindhoven University of Technology, 5600 MB Eindhoven, The Netherlands}
\email{l.m.c.janssen@tue.nl}

\date{\today}

\begin{abstract}
Soft particles such as microgels can undergo significant and anisotropic deformations when adsorbed to a liquid interface. This, in turn, leads to a complex phase behavior upon compression. To date, experimental efforts have predominantly provided phenomenological links between microgel structure and resulting interfacial behavior, while simulations have not been entirely successful in reproducing experiments or predicting the minimal requirements for a desired phase behavior. Here we develop a multiscale framework to rationally link the molecular particle architecture to the resulting interfacial morphology and, ultimately, to the collective interfacial phase behavior. To this end, we investigate interfacial morphologies of different poly(N-isopropylacrylamide) particle systems using phase contrast atomic force microscopy and correlate the distinct interfacial morphology with their bulk molecular architecture. We subsequently introduce a new coarse-grained simulation method that uses augmented potentials to translate this interfacial morphology into the resulting phase behavior upon compression. The main novelty in this method is the possibility to efficiently encode multibody interactions, the effects of which are key in distinguishing between heterostructural (anisotropic collapse) and isostructural (isotropic collapse) phase transitions. Our unifying approach allows us to resolve existing discrepancies between experiments and simulations. Notably, we demonstrate the first accurate in silico account of the two-dimensional isostructural transition, which is frequently found in experiment but elusive in simulation. In addition, we provide the first experimental demonstration of a heterostructural transition to a chain phase in a single-component system, which has been theoretically predicted decades ago. Overall, our multiscale framework provides a bridge between physicochemical soft-particle characteristics at the molecular- and nanoscale and the collective self-assembly phenomenology at the macroscale, paving the way towards novel materials with on-demand interfacial behavior.
\end{abstract}


\maketitle 
\section*{Introduction}
Adsorption of colloidal particles to liquid interfaces is ubiquitous and relevant in both fundamental science and technological applications~\cite{McGorty2010,Binks2002,Pieranski1980}. For example, in functional soft matter applications, the presence of adsorbed particles imparts the kinetic stability of emulsions~\cite{Liu2012,Destribats2012}, liquid marbles~\cite{Aussillous2001,Binks2006} and foams~\cite{Binks2005,Gonzenbach2006}. Additionally, colloidal particles can serve as masks to obtain complex nanoscale surface patterns~\cite{grillo2020,Fernandez-Rodriguez2018,Tang2018,Goerlitzer2020}.  
Furthermore, since liquid interfaces confine the particles on two dimensions, they serve as ideal templates for fundamental studies of colloidal interactions, crystallization and self-assembly~\cite{McGorty2010,Pieranski1980,Kaz2011}.

At liquid interfaces, colloidal particles are able to self-assemble into ordered two-dimensional (2D) crystals~\cite{Vogel2015}. Depending on the balance between attractive capillary and van der Waals forces and repulsive dipole and electrostatic forces~\cite{Pieranski1980}, monodisperse spherical particles typically form hexagonal close-packed or non-close-packed structures~\cite{Aveyard2000}. 
In recent years, the interfacial behavior of soft particle systems at liquid interfaces has been attracting attention~\cite{Rey2020}.
In contrast to their rigid analogues, soft particles can deform significantly and anisotropically under the influence of surface tension~\cite{Style2015}, and may assume a characteristic core-corona morphology~\cite{Geisel2012,Zielinska2016, Rauh2017,Camerin2019}  to cover more interfacial area than assumed from their bulk diameter. This change in morphology leads to a more complex interfacial phase behavior. 
At low surface pressure $\Pi$ (corresponding to a large available area per particle $A_P$), soft particles typically assemble into a hexagonal non-close-packed configuration in which the particle coronae are in  contact~\cite{ReySM2016,Rey2020,Rauh2017,ReySM17,Picard2017,Scheidegger2017,El-Tawargy2018,Tang2018,Vasudevan2018,Scotti2019}. Upon compression to decrease the available area at the interface, different phase behaviors may be observed: either the hexagonal phase remains intact, with a lattice constant that decreases continuously (continuous transition) or discontinuously (isostructural transition) upon compression, or a symmetry-breaking transition occurs toward a new phase without hexagonal order. We refer to the latter as a heterostructural transition. 

Despite their broad relevance, the 2D phase behavior of soft particles is still confounded by significant discrepancies between theory and experiment. Indeed, the isostructural transition is experimentally among the most commonly observed behaviors~\cite{Vogel2012,ReySM2016,Rauh2017,ReySM17,Picard2017,Scheidegger2017,El-Tawargy2018,Tang2018,Vasudevan2018}, 
yet in computer simulations it has proven notoriously difficult to model a discontinuous transition between a non-close-packed and close-packed hexagonal phase in a monodisperse system. Conversely, 2D heterostructural transitions toward e.g.\ a chain phase~\cite{Jagla1998,Glaser2007,Du2019,Somerville2020,Malescio2003}, clusters~\cite{Fornleitner2010}, honeycomb~\cite{Pattabhiraman2017}, or quasicrystalline~\cite{Dotera2014, Pattabhiraman2017} phases can be simulated via relatively simple, Jagla pair potentials~\cite{Jagla1998,Jagla1999,Jagla1999a}. 
However, to date, no experimental single-component system has exhibited such heterostructural behavior, even though the synthesized particles should, at first sight, behave according to Jagla. These apparent incongruities in experiment and simulation underline the need for a more rigorous understanding of the relation between molecular soft-particle properties, interfacial morphology and macroscopic interfacial phase behavior. 

In this work we resolve these discrepancies by establishing a unifying multiscale-based framework for the 2D phase behavior of monodisperse soft particles. We combine experiments and a new simulation approach to rationally connect the collective phase behavior on the macroscale to the single-particle morphology on the nanoscale. Our framework recognizes that the emergent phase behavior ultimately stems from (anisotropic) two- and many-body interactions among particles; these interactions, in turn, must follow from the molecular architecture of each individual particle. Thus, by carefully constructing the morphology that the microgels assume at the liquid interface via their molecular architecture, we are able to controllably achieve continuous, isostructural, and heterostructural 2D phase transitions. 

\section*{Results and discussion}

In the following we demonstrate how to control the 2D phase behavior of single-component soft particle systems. 
First, let us briefly recall the distinctive features of the different phase regimes we target in this work: the \textit{continuous} regime corresponds to a continuous decrease in lattice spacing upon compression, whilst preserving the overall hexagonal lattice structure. The \textit{heterostructural} regime is characterized by a change in the symmetry upon compression, for example from hexagonal to chain phases. Lastly, the \textit{isostructural} case is characterized by a discontinuous transition that proceeds via hexagonal clusters of close-packed particles that nucleate within a non-close-packed hexagonal phase.
As detailed below, our simulation approach is able to capture all different phase transition regimes within one coherent framework.
In combination with experiments, this approach allows us to recognize common features among particles from the same regime and to provide hypotheses connecting the interfacial morphologies with the macroscopic phase behavior. These newly identified connections finally enable us to tailor the particle architecture to achieve a required interfacial morphology, and thus, the desired phase behavior.

\subsection*{From molecular architecture to interfacial morphology}
Our proof-of-principle is established for three different soft particle architectures, each representative of a different 2D phase behavior and chosen with the support of our theoretical insights described below. 
We use poly(N-isopropylacrylamide) (PNIPAM) microgels with ($N$,$N'$-methylenebisacrylamide) (BIS) as crosslinker as a widely-used model system with well-established synthetic protocols~\cite{Pelton2000}.
As the first particle system we synthesize a microgel with very low crosslinking density ($1\mpc$ BIS), as used in our previous work~\cite{ReySM17}. 
The second and third particle systems are silica-PNIPAM core-shell structures~\cite{Tang2018}, providing two distinct length scales, which are given by the hard silica core and the soft PNIPAM shell, respectively.
We capitalize on the known difference in reactivity of the crosslinker BIS compared to the NIPAM monomer~\cite{Kroger2017,Wu1994}, which leads to a crosslinking gradient from the center towards the periphery of microgels~\cite{Fernandez-Barbero2002}, and vary the uniformity of the crosslinker distribution, while keeping the overall concentration similar ($5\mpc$ BIS). In the second particle system, we create a non-uniform crosslinking density through a batch synthesis process of the PNIPAM shell. In this case, the higher reactivity of the crosslinker is expected to lead to a decrease crosslinking density towards the periphery of the microgel shell.
In the third particle system, we pursue the opposite strategy and ensure a uniform crosslinking density throughout the shell by continuous feeding of monomer and crosslinker.~\cite{Tang2018,Still2013,Kyrey2019}.
Note that with the exception of the second particle system, all particle systems have been synthesized and used in literature before. Here, our aim is to re-examine their interfacial behavior in the framework of our simulation method, and to rationalize the resulting phase transitions on the basis of their interfacial morphology and the (an)isotropy of the pressure-induced collapse.  

\begin{figure*}
\centering
\includegraphics[width=\textwidth]{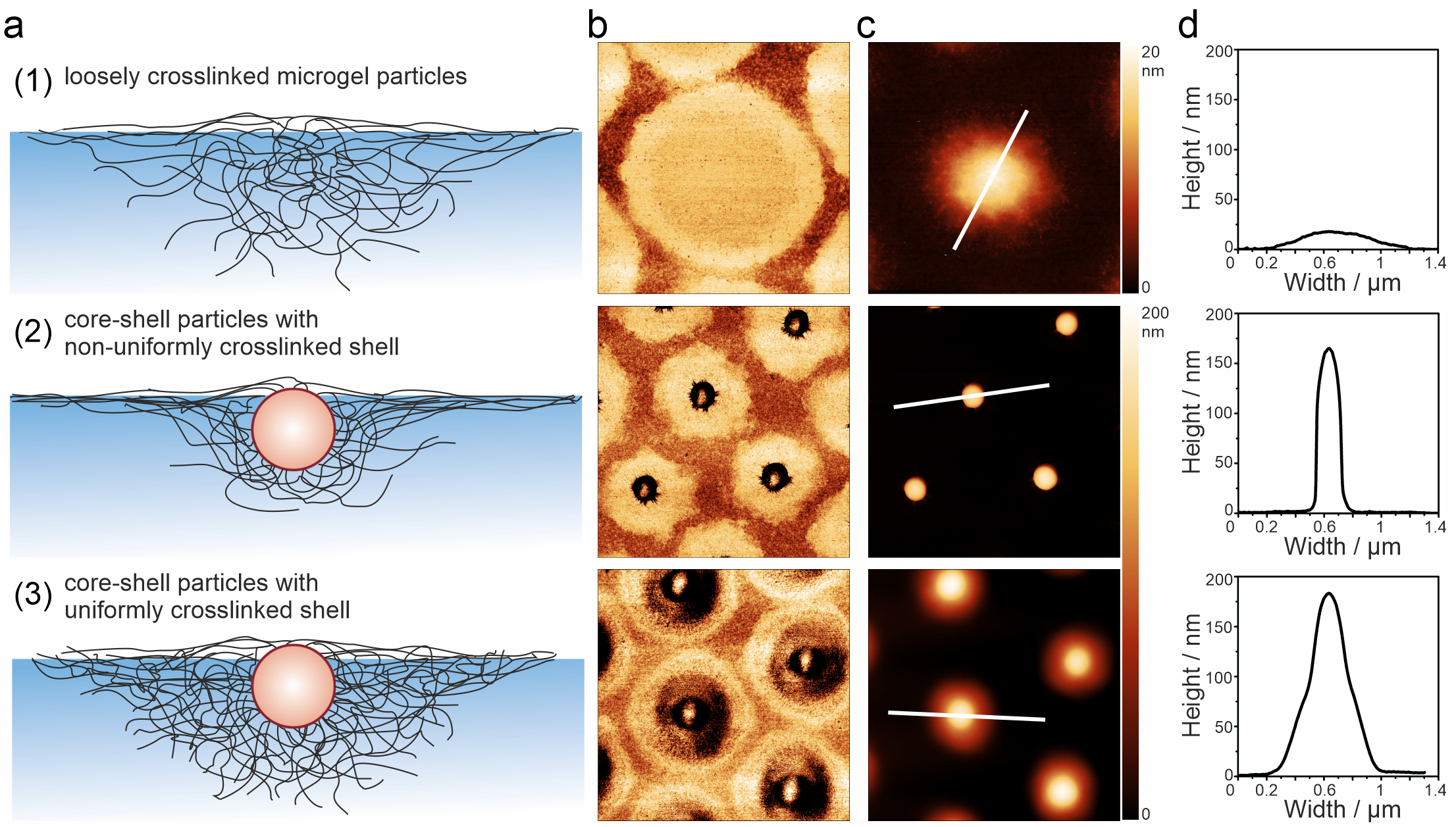}
\caption{The molecular architecture of soft microgel systems controls their interfacial morphologies. a) Schematic illustration of the hypothesized interfacial morphology. b-d) Corresponding AFM phase (b) and height (c) images.
The cross-section (d) is measured along the white lines of (c) , highlighting the difference in their interfacial morphology. Images 2x2 $\mu m^2$.}
\label{fig:structures}
\end{figure*}

To first connect the molecular architecture of the individual particle systems with their interfacial morphology, we assemble the particles at the air/water interface without compression (surface pressure $\Pi=0$) and deposit them onto a solid substrate~\cite{Rey2020}. We then characterize their morphology in the dry state using atomic force microscopy (AFM) (Fig.~\ref{fig:structures}). Although drying will cause changes to the overall microgel dimensions, we assume that the morphology of the particle at the interface remains intact. We and others have previously shown that both the interfacial morphology and the hexagonal symmetry of the assembly phases are effectively preserved after transfer to a solid substrate and that they resemble the shape of microgels imaged directly at the interface~\cite{Rey2020,ReySM2016,Rauh2017,Scheidegger2017}. This observation indicates that capillary forces are unlikely to change the microgel morphology, which is further supported by the known affinity of microgels to silica surfaces~\cite{Rey2020a}. It is known that microgels deform at the air/water interface and assume a characteristic core-corona morphology, where the core consists of the more crosslinked central region of the microgel, while the corona forms as an extremely thin polymer layer, presumably by dangling polymer chains stretching out from the microgel periphery.~\cite{Geisel2012,Rauh2017,Zielinska2016,Camerin2019,Rey2020}
Because of its very thin nature, this corona is often not visible in SEM of AFM height images, but can be clearly recognised by its contrast in AFM phase images.~\cite{Rey2020}

We observe that all three particle systems expand at the liquid interface under the effect of surface tension and assume a core-corona morphology visible in the AFM height and phase images (Figure~\ref{fig:structures}b,c). Although their composition and bulk dimensions are comparable (Fig.~S1a), the three different particle systems exhibit substantial differences in their interfacial morphology. The AFM height image of the loosely crosslinked microgels (Fig.~\ref{fig:structures}, (1)) without the silica core displays a broad, flattened profile with a continuous and regular thinning from the center, resulting from the particle's high degree of deformability (Fig. 1,c top). In addition, the AFM phase image clearly reveals a corona surrounding the center region, which is too thin to be seen in the AFM height image (Fig. 1c, top). We therefore hypothesize that the shape of the microgels at the liquid interface will be similarly three-dimensional with a continuous decrease in height towards the periphery as seen in the AFM height image.
For the core-shell particles with non-uniform crosslinked shells (Fig.~\ref{fig:structures}, (2)) we observe a well-defined core in the AFM height image with an abrupt transition to the corona, which is only visible in the AFM phase image due to its extremely thin nature. We assume that the same morphology is present at the liquid interface, i.e.\ these particles exhibit a very flat and extended two-dimensional corona. This morphology reflects the ability of the loosely connected polymer chains at the outer periphery of the microgel shell to extend at the interface to minimize surface tension. 
For the core-shell particles with uniform crosslinking density (Fig.~\ref{fig:structures}, (3)) we observe a more continuous density profile from center to periphery. In the AFM height image, a deformed polymeric shell surrounding the silica core  is visible. The AFM phase image reveals the presence of an additional, extremely thin corona. The uniform crosslinker distribution in this sample prevents the polymer chains to fully extend at the interface. As a consequence, the shell retains an appreciable volume and only forms a much smaller corona at the periphery, compared to sample (2).

\subsection*{Augmented-potential simulation model}
\begin{figure*}[!]
\centering
\includegraphics[width=0.75\textwidth]{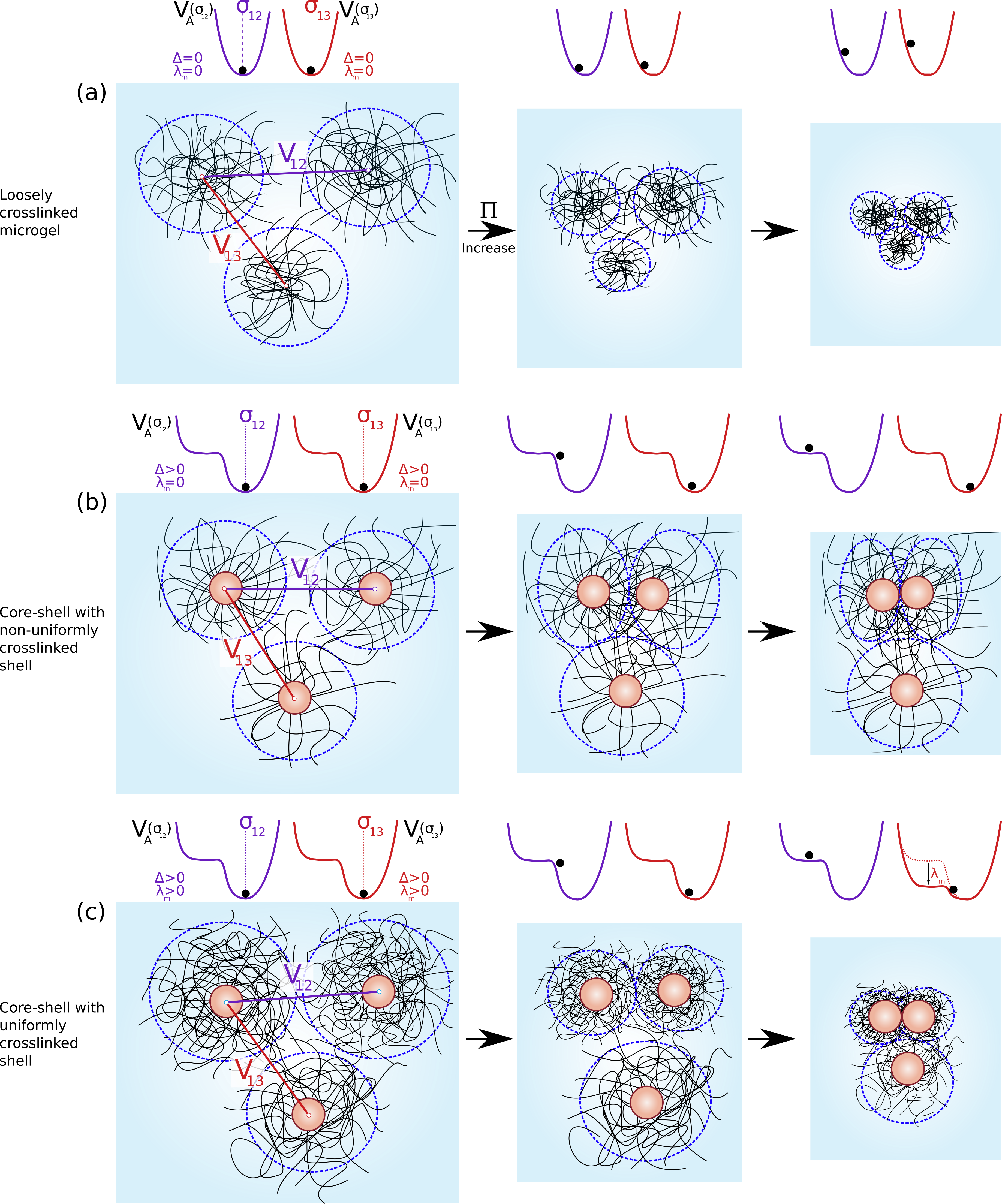}
\caption{Augmented potential-based simulation scheme used to capture the different interfacial morphologies and simulate the resulting phase behavior. (a)-(c) schematic illustration of the response of the different particle systems to compression. The average shape is indicated by dashed blue lines. To model the effect of the compression on the particle shape and to capture all the phase transition regimes, we use augmented potentials as a function of the interaction distance $\sigma_{ij}$, as indicated on top of the individual schematics. a) The shape of loosely crosslinked microgel particles is represented by setting a barrierless ($\Delta=0$) augmented potential (eq.~\ref{eq:dw}) to mimic the fact that such particles always push to expand to their equilibrium size, and by excluding many-body effects ($\lambda_m=0$). b) The shape of core-shell particles with a non-uniformly crosslinked shell is represented using $\Delta>0$ in the augmented potential defined in eq.~\ref{eq:dw}, thus creating a metastable state that corresponds to core-core contacts, that can be reached for sufficiently high surface pressure. c) Uniformly crosslinked shells are modelled by including multibody effects. In contrast to loosely crosslinked shells, this particle system is only allowed to collapse isotropically; this effectively reduces the repulsion among all neighboring particles in a strongly correlated manner. We simulate this multibody effect by setting $\lambda_m>0$. See the Methods section for more details. 
}
\label{fig:aug_phases}
\end{figure*}


To  model the phase behaviors of the three different systems in silico, we require a simulation method that can account for all the relevant soft particle properties, but that at the same time is sufficiently computationally efficient to simulate a large collective of particles.\cite{Harrer2019a,Camerin2019,Ninarello2019,Rovigatti2019,Rovigatti_2017,Gnan2017,Choudhury2020,Bushuev2020}
Although coarse-grained molecular dynamics (MD) simulations with simple pair potentials are already capable of predicting continuous and heterostructural transitions,\cite{Jagla1998,Jagla1999,Jagla1999a} the 2D isostructural phase transition has not yet been captured by conventional MD methods. Here we introduce a novel, more advanced simulation method based on augmented variables, allowing us to explicitly account for a partial or full collapse of the coronae, as well as for isotropic and anisotropic corona deformations, depending on the specific particle morphology and surface pressure. This augmented approach can effectively capture many-body effects without the need of adding many-body potentials.\cite{stillinger1985computer,Tersoff1989,Allahyarov1998,Brenner2002,Ciarella2018} As argued below, this simulation methodology not only reproduces the well-established continuous and heterostructural transitions, but it also naturally captures, for the first time, the 2D isostructural transition in single-component systems.

To accurately reflect the essential features of each particle system in  simulation, we recognize that (i) the effective particle size at a given surface pressure follows from the particle softness, (ii) the softness effectively follows from the crosslinking gradient in the polymer network, and (iii) the crosslinking gradient also sets the degree of isotropy; that is, whereas loose polymer chains at the interface may behave independently from each other to yield an anisotropic (2D) corona deformation, more crosslinked chains must behave in a more concomitant, and therefore more isotropic, manner. 
We account for these effects by 
letting the typical equilibrium interaction ranges $\sigma_{ij}$ between particles $i$ and $j$ evolve as augmented, dynamical variables that can vary between extended and collapsed states depending on the local particle environment. We also note that within our approach, and similar to standard coarse-grained simulations of binary mixtures~\cite{Kob1994}, the size of an individual particle $i$ is not an explicit model parameter, but rather follows implicitly from the pair interaction parameter $\sigma_{ij}$. Importantly, this also allows us to describe \textit{anisotropic} particle deformations in the case of non-uniformly crosslinked shells, namely by letting $\sigma_{ij}$ and $\sigma_{ij'}$ differ among two neighboring particles $j$ and $j'$. For visual purposes, we will assume that the effective particle radius $r_i$ is given by $r_i = r_{i,\textrm{min}} = \min_{j} (\sigma_{ij}/2)$, where the index $j$ runs over all possible neighbors of particle $i$. 

As detailed in the Methods section and further below, we can capture all three experimental particle systems within the same simulation framework and transition between them by tuning only two parameters: an energy barrier $\Delta$ that sets the presence of distinct internal length scales in the particle morphology, and a parameter $\lambda_m$ that sets the degree of isotropy in particle compression, and therefore the strength of multibody interactions.
For all bare pair interactions, we use a generalized Lennard-Jones potential. We choose the potential parameters such that the preferred particle sizes in  simulation are consistent with experiment, as explained in more detail in the Methods section. 
In Fig.~\ref{fig:aug_phases} we sketch our augmented potential and the corresponding physical scenarios; the augmented potentials are also discussed in more detail in Fig.~\ref{fig:pot}(a). 
\subsection*{Collective interfacial phase behavior} 

We now analyze the collective phase behavior of the three distinct soft particle systems confined at a 2D liquid interface. Experimentally, we characterize the systems at the air/water interface using the simultaneous compression and deposition method~\cite{ReySM2016, Rey2020}, and employ ex situ scanning electron microscopy (SEM) combined with a custom-written image analysis software (details in Sup. Info) to establish the different structural phases. The surface pressure - area per particle diagram can be found in Figure S1b.
We compare the observed interfacial phase behavior to the results of our augmented potential model and discuss how the phase behavior can be ultimately rationalized from the molecular architecture via the differences in interfacial morphology. 

\subsubsection*{Loosely crosslinked particles: Continuous regime} 
\begin{figure*}[!t]
\includegraphics[width=\textwidth]{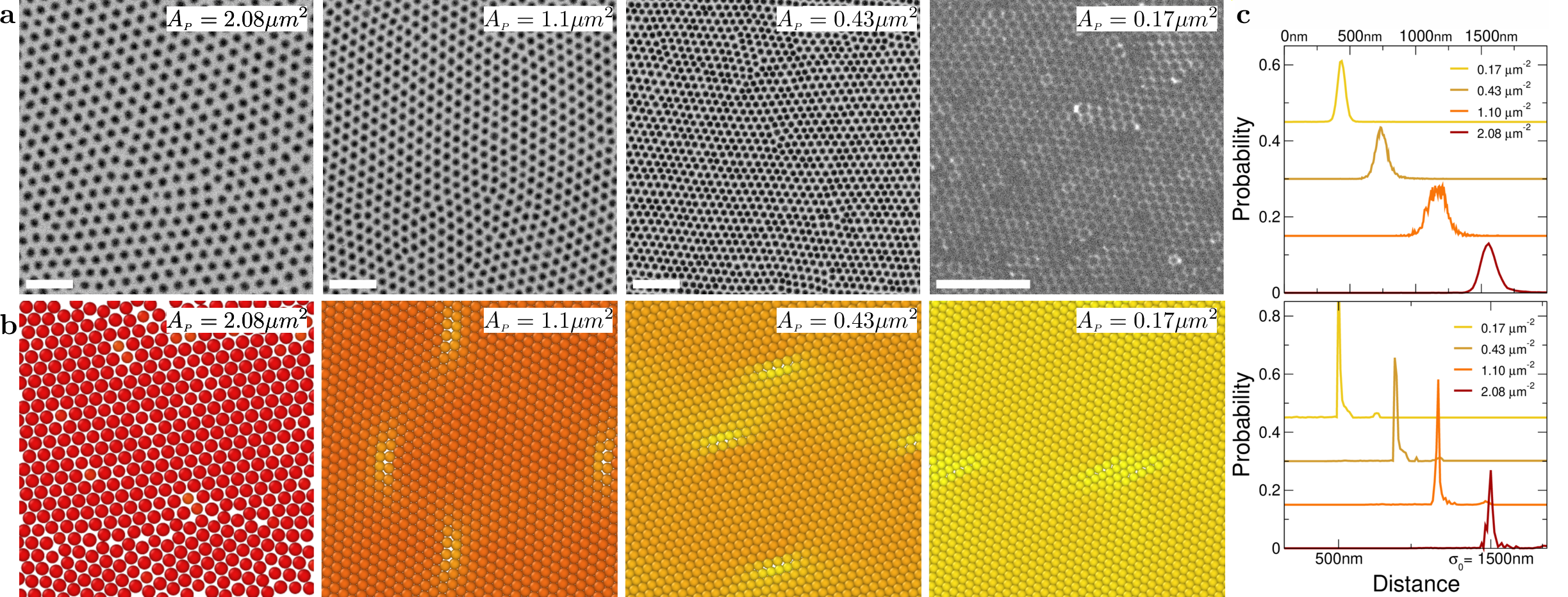}
\caption{Microgels with a very low crosslinking density behave according to the \textit{continuous} regime. Comparison between experimentally observed phases (a) and simulations with augmented potentials ($\Delta=\lambda_m=0$) (b) for microgel particles showing a \textit{continuous} transition from a hexagonal non-close-packed to a close-packed phase. The area per particle $A_P$ decreases from left to right. The experimental scale bar is 5 $\mu m$, while simulated particles are red in their expanded size and they follow a gradient toward yellow during compression. 
The microgel particles gradually shrink upon compression while preserving their hexagonal lattice.  A statistical analysis of the distributions of the nearest-neighbor distances upon compression is shown for both experiment and simulation (c). Datasets are progressively shifted vertically by $0.15$ for clarity.}
\label{fig:reg1}
\end{figure*}

Upon compressing microgels with a low crosslinking density (particle system (1)), an interfacial behavior classified as the \textit{continuous} regime is found.
Indeed, Fig.~\ref{fig:reg1} shows the distinctive uniform shrinkage of microgels upon compression, giving rise to a continuous transition from non-close packed to close-packed hexagonal order. 

In our experiments, the softness is associated with the characteristic semi-spherical morphology of such particles, visible in Fig.~\ref{fig:structures} (1). 
This quasi-three-dimensional structure causes a rapid increase of overlap volume between the coronae of two neighboring microgel particles upon compression at the interface, i.e.\ with increasing surface pressure~\cite{ReyJACS17}. 
Assuming that the net (repulsive) interaction energy is proportional to the overlap volume of the coronae, and that the overlap volume between quasi-three-dimensional spheres grows non-linearly with the particle-particle distance, it follows that the repulsion rapidly increases with compression.
This effectively forces the particles to simply pack closer together in a continuous manner.  

The continuous regime also naturally appears in our augmented simulation approach with a single equilibrium distance corresponding to two expanded microgels in contact (Figure 2a). 
We recall that the continuous regime can also be captured by simpler~\cite{Jagla1998,Jagla1999} or more refined~\cite{Camerin2020} coarse-grained models using convex repulsion potentials.
In the augmented scheme, this convexity appears in both the particle-particle interaction and the augmented potential; in fact, the augmented potential $V_A$ which we use to describe this microgel system is always convex if we set $\Delta=0$ (see Methods).
As shown in Fig.~\ref{fig:reg1}, our simulations closely reproduce the experimental phase behavior: a decrease in the area per particle fully preserves the hexagonal lattice but monotonically decreases the lattice constant. 

The statistical distribution of nearest-neighbor distances further underscores that the continuous transition is correctly captured both in experiments and simulation (Fig.~\ref{fig:reg1},c).
It can be seen that the typical interparticle distance, and thus the lattice spacing, decreases continuously upon compression. This behavior is consistent with the fact that such particles lack a defined core and therefore do not possess an additional internal length scale, rendering our augmented potential with a single length scale ($\Delta=0$) suitable for the modeling. 

\subsubsection*{Core-shell particles with non-uniformly crosslinked shell: Heterostructural regime}

\begin{figure*}[!t]
\includegraphics[width=\textwidth]{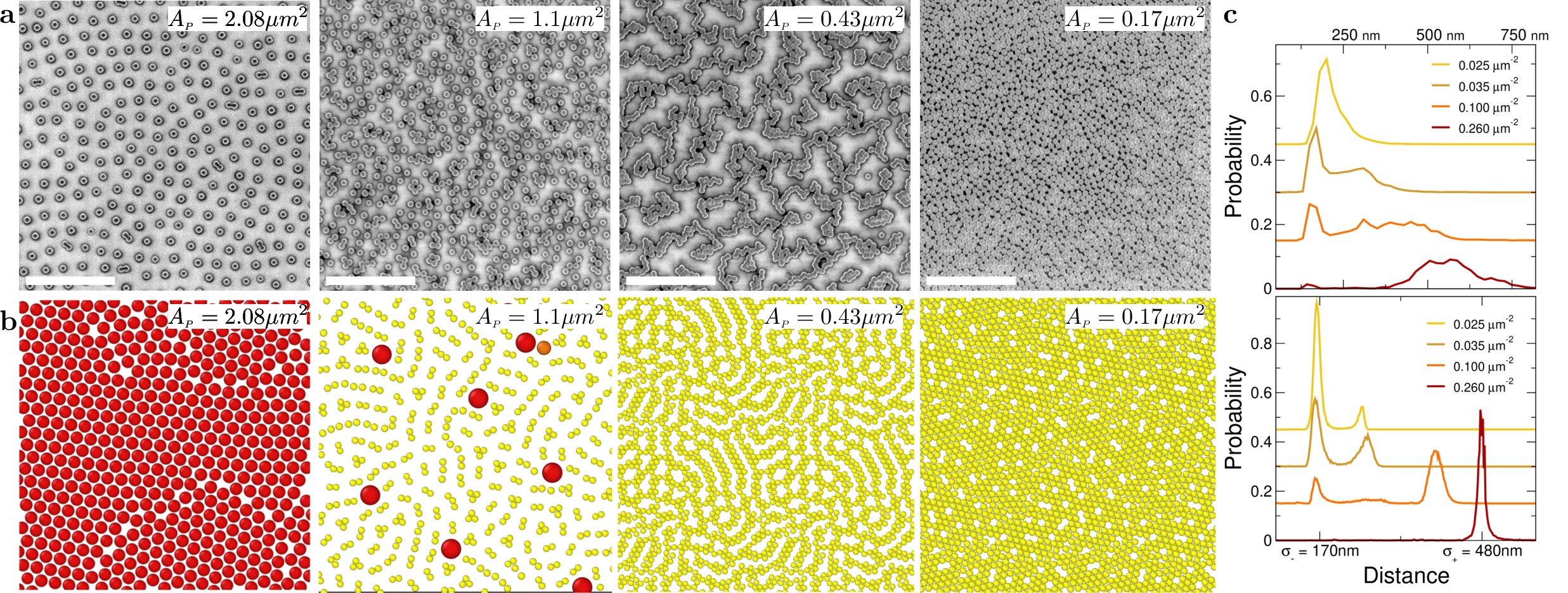}
\caption{Core-shell particles with non-uniform crosslinker distribution in the shell behave according to the \textit{heterostructural} regime. Comparison between experimental phases (a) and simulations with augmented potentials ($\Delta>0$,$\lambda_m=0$) (b) showing a \textit{heterostructural} transition from a hexagonal non-close-packed to a chain to a hexagonal close-packed phase. The area per particle $A_P$ decreases from left to right. The experimental scale bar is 2.5 $\mu m$, while simulated particles are red in their expanded size and they follow a gradient toward yellow while collapsing. Signatures of the heterostructural transition are the chains of collapsed particles forming at intermediate compression.  In (c) we report the distribution of the nearest-neighbor distance upon compression. Datasets are progressively shifted vertically by $0.15$ for clarity.
}
\label{fig:reg2}
\end{figure*}


Figure~\ref{fig:reg2} shows the experimental and numerical results for the heterostructural transition regime, which exhibits phase transitions from a hexagonal non-close-packed phase to a characteristic, anisotropic chain phase, and finally to a hexagonal close-packed structure upon compression. The transition is further confirmed in the nearest-neighbor distance distribution, which reveals the chain phase by a coexistence of collapsed and more extended particle pairs. Our experimental data with non-uniformly crosslinked particles (particle system (2)) constitute the first single-component system exhibiting such a heterostructural transition. Previously, the only observed chain phases occurred in soft binary  mixtures with similar assumed interfacial morphologies~\cite{ReyJACS17,Rey2018,Rey2018a}.


The ability to achieve this phase transition experimentally stems from the theoretical insights gained from our simulations (Fig.~\ref{fig:aug_phases}), which demand not only two distinct internal particle length scales, but also an anisotropic type of interaction, such that $\sigma_{ij}$ and $\sigma_{ij'}$ can be distinct for a given particle $i$ with neighbors $j$ and $j'$. We can realize both of these requirements in core-shell particles with a non-uniform crosslinker distribution in the shell, and, thus, more loosely connected polymer chains in the periphery.
To understand the rationale behind this, let us first recall from Fig.\ \ref{fig:structures}(2) that these particles assume an pronounced core-corona morphology with a well-defined hard core and a thin, quasi-two-dimensional corona when adsorbed at the liquid interface. The clear distinction between core and corona naturally satisfies the two-length-scales prerequisite. The second condition requires the corona to be anisotropically compressible at the interface in any direction. Physically, this enables the system to leverage a trade-off in repulsive corona overlaps upon compression: rather than letting \textit{all} pairs of particles undergo a partial, uniform overlap (or collapse) of their coronae, the system may minimize its total free energy by avoiding any overlap among neighboring particles in one direction, at the cost of fully overlapping in another. This leads to a chain phase in which the intrachain interactions correspond to core-core contacts, while the interactions between adjacent chains are governed by extended corona-corona contacts. On the molecular level, such anisotropic corona deformations can be achieved by loosely connected polymer chains that can interact with neighboring particles independently from one another. Thus, the ideal corona should consist of many individual long polymer chains with as little chemical or physical crosslinking points as possible. The experimental results in Fig.~\ref{fig:reg2} suggest that our chosen particle morphology realizes this scenario, indicating that in the batch synthesis method, the crosslinker is indeed mainly integrated around the inner region of the polymer shell. Finally, with these findings, we can also infer why previous experimental studies on  core-shell particles have not found heterostructural transitions, namely due to a too high crosslinking density in the periphery that prevents anisotropic corona interactions. 

In contrast to experiments, the heterostructural regime is less difficult to find in silico and has been theoretically predicted decades ago by particles interacting via soft repulsive, Jagla-type potentials.~\cite{Jagla1998,Jagla1999,Jagla1999a}  Indeed, since many-body correlation effects can be ignored due to the flexibility of the coronae in all directions, it suffices to let particles interact through a relatively simple pairwise potential. This interaction, however, must include two different length scales to account for the distinct core-core and corona-corona contacts. For Jagla-like potentials used in standard MD simulations, one has to be careful and additionally destabilize the continuous regime by invoking a non-convex potential (or $g>1$ in Jagla's terminology)~\cite{ReyJACS17}.
We note that this Jagla-based perspective can also be united with the observed interfacial morphology (Fig.\ \ref{fig:structures}(2)).
Again assuming that the repulsive interaction energy scales with the overlap volume, we expect a near-linear ramp potential for two-dimensional corona morphologies, as opposed to the more semi-spherical conformations discussed in the previous section~\cite{ReyJACS17}.
In our simulation scheme, we rely on the augmented potential $V_A$ by setting $\Delta>0$ to incorporate the two required length scales, without a stringent convexity requirement on the bare interaction potential (Figure 2b).
An additional advantage of this augmented ensemble approach over the Jagla approach is that particles can also be attractive (see Methods and eq.~\ref{eq:lj}) to account for e.g.\ low-compression capillary effects~\cite{Ciach2017}.
 

\subsubsection*{Core-shell particles with uniformly crosslinked shell: Isostructural regime}

\begin{figure*}[!]
\includegraphics[width=\textwidth]{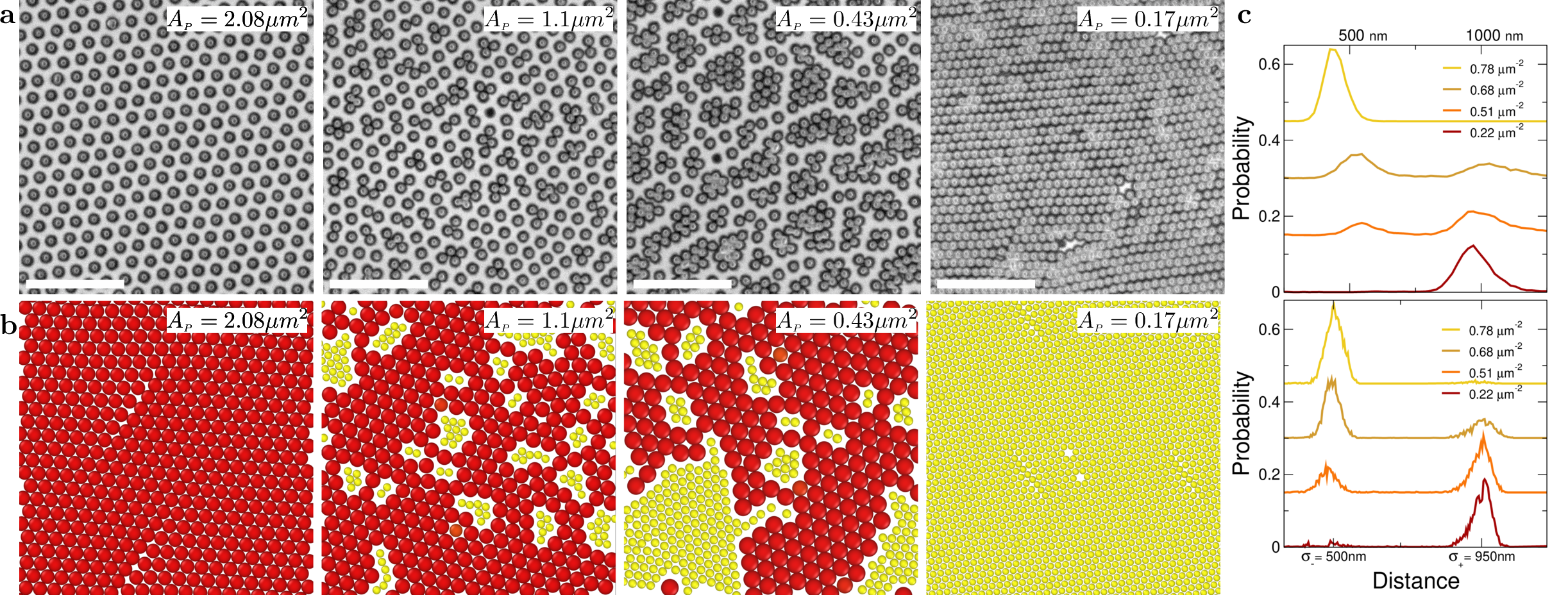}
\caption{Core-shell particles with a uniformly crosslinked shell behave according to the \textit{isostructural} regime. Comparison between experimental phases (a) and simulations with augmented potentials ($\Delta>0$,$\lambda_m>0$) (b) show a transition from hexagonal non-close-packed to hexagonal close-packed. The are per particle $A_P$ decreases from left to right. The experimental scale bar is 5 $\mu m$, while simulated particles are red in their expanded size and they follow a gradient toward yellow while collapsing. A characteristic of the isostructural transition is the coexistence of clusters of collapsed particles and expanded particles at intermediate surface pressure. (c) shows the distribution of the nearest-neighbor distance upon compression. Datasets are progressively shifted vertically by $0.15$ for clarity.
}
\label{fig:reg3}
\end{figure*}

The third phase regime concerns the isostructural phase transition of soft particles, which has been commonly observed in experiments~\cite{ReySM2016,ReySM17,Picard2017,Scheidegger2017,Rauh2017,Tang2018,Vasudevan2018,El-Tawargy2018}, but which thus far could only be
predicted by simulation~\cite{Bolhuis1997} and theory~\cite{Denton1997,Hemmer2001,Navascues2016} for very short shells that do not match the experiments, and for three-dimensional bulk phases. 
In Fig.~\ref{fig:reg3} we present the first in silico demonstration of the 2D isostructural phase transition obtained from our augmented simulations, combined with the corresponding experimental results for core-shell particles with a uniform crosslinking distribution in the shell.  
The distinctive growth of close-packed hexagonal clusters is evident at intermediate compression for both simulations and experiments. 
Furthermore the histograms in Fig.~\ref{fig:reg3}c show a bimodal distribution of interparticle distances, corresponding to either expanded or collapsed pairs.

To understand the physical origins of the discontinuous isostructural transition, we first recognize that particles must exhibit a sharp and isotropic collapse upon compression. 
Naturally, the sharp distinction between a collapsed and extended state also requires two internal particle length scales. Core-shell particles with a relatively densely and uniformly crosslinked shell (particle system (3)) are a typical example of such  particles. 
Figure~\ref{fig:structures} (3) shows their interfacial morphology, clearly satisfying the prerequisite of having two distinct length scales.
Additionally, with the continuous-feeding synthesis method, the resulting polymer shells contain more uniformly distributed crosslinking points, which is consistent with the continuous decrease in height from the core to the periphery, as seen in Fig.~\ref{fig:structures} (3). 
In this molecular architecture, the chains are much more constrained by the presence of crosslinkers across the entire volume of the shell, which reduces both the length of free dangling chains at the periphery, and the freedom of the chains to move independently from each other. Upon compression, such uniformly crosslinked particles are therefore forced to distribute the stress across the entire shell.
This in turn, prohibits the formation of a chain-like phase, since the corona is no longer able to collapse in one direction whilst simultaneously remaining extended in another. Instead, we postulate that the collapse of any pair of particles now imposes a shrinkage of the coronae in all directions, and thus facilitates collapse events with other neighboring particles. 

From the modeling side, the isostructural regime has been notoriously difficult to capture in standard MD simulations.
For example, the Jagla potential only predicts a continuous transition between the two hexagonal phases and not an isostructural phase transition, regardless of the chosen pair potential~\cite{Jagla1999a,ReyJACS17,Rey2018a}. From our insights, we can attribute these earlier discrepancies to the fact that pairwise potentials in conventional MD models impose interactions that are fixed a priori, so that all particle pairs always interact in the same way at a given interparticle distance.
However, in reality, an \textit{isotropic} particle collapse induced by one neighbor will concomitantly affect the interactions with all other neighboring particles. Indeed, it follows from the interconnected polymer shell that the collapse will lead to a significant change in particle conformation in all possible directions. To adequately model the isostructural regime, it is therefore crucial to incorporate local many-body effects into the simulations, such that any isotropic particle deformation is immediately translated to a change in the net pair potentials with neighboring particles. 

Keeping these considerations in mind, we can rationally adopt our augmented simulation scheme to describe the isostructural regime. Specifically, we  account for distinct collapsed and extended states by setting $\Delta>0$ in the augmented potential $V_A$; for the required local many-body effects we impose one additional constraint by setting $\lambda_m>0$ (Figure 2c). To understand the latter, 
let us first recall that in the heterostructural regime every particle $i$ is allowed to interact independently (i.e.\ with a different $\sigma_{ij}$) with each of its neighbors $j$. 
A pair $ij$ can then either be in an expanded state $\sigma_{ij}\sim\sigma_+$ or, if compressed, in a collapsed state $\sigma_{ij}\sim\sigma_-$.
In the isostructural regime, we assume that once a particle $i$ collapses with one of its neighbors 
such that the minimum particle radius becomes $r_{i,\textrm{min}} <\lambda_m \sigma_-/2$, 
it will also immediately collapse with all the other neighbors $j$ that are already involved in a collapse elsewhere (i.e.\ with $r_{j,\textrm{min}} <\lambda_m \sigma_-/2$).
Subsequently, we then impose that $\sigma_{ij}=r_{i,\textrm{min}}+r_{j,\textrm{min}}=\sigma_-$. This mimics the effect that each collapse after the first is facilitated by the presence of other collapsed particles nearby, as we sketch in the bottom right panel of Fig.~\ref{fig:aug_phases}. 
When $\lambda_m=0$ this effect is suppressed and each pair collapses individually.
When $\lambda_m\sim 1$ this multi-body interaction becomes relevant because the resulting $\lambda_m \sigma_-/2$ is an available state.
Note that this also effectively suppresses the formation of chain-like structures, since a stable chain phase would require that $\sigma_{ij}=\sigma_+$ for interactions between particles in adjacent chains (also see Methods and Fig.~\ref{fig:rule}). 

\section*{Conclusion}
In this work, we put forward a new multiscale-based perspective of soft particle behavior at liquid interfaces, allowing us to rationally control and design collective interfacial phase behavior by tailoring the interfacial morphology via the molecular architecture of soft particles. 
To asses our design rules, we tailor three archetypal soft particles that correctly exhibit the three characteristic phase behaviors of continuous, heterostructural, and isostructural phase transitions.
Our experimental observations are supported by a newly developed augmented simulation model which is based on similar design rules, and, for the first time, is able to capture all experimentally observed phase transitions, including the elusive 2D isostructural transition, in a unifying framework. 

The general design rules established in this work can be summarized as follows: 

(i) The \textit{continuous} transition is characterized by a single length scale and a repulsion energy that increases non-linearly (convexly) with particle-particle distance.
Soft particles without a hard core and a loosely crosslinked structure are the perfect prototype for this continuous regime, since their quasi-semi-spherical interfacial morphology naturally yields a net convex interaction potential.

(ii) For a \textit{heterostructural} phase transition, where non-close-packed and close-packed hexagonal phases are separated by other ordered phases such as chains, a core-shell particle with two distinct internal length scales is required. 
Additionally, the corona needs to be anisotropically compressible to allow particles to collapse in one direction, i.e.\ within a chain, while avoiding overlap in another, i.e.\ between two adjacent chains. 
We demonstrate that this phase behavior can be experimentally realized for core-shell particles with a non-uniformly crosslinked shell, where the periphery consists of long dangling chains with minimal crosslinking.

(iii) Lastly, the \textit{isostructural} regime, which is characterized by a discontinuous transition from a non-close-packed to close-packed hexagonal phase, also requires two internal particle length scales. 
However, as opposed to the heterostructural case, the particles must collapse isotropically to distribute the pressure load in every direction. Experimentally, this behavior is favored by a more uniformly crosslinked shell that forces the polymer chains to compress in a concomitant manner.
Theoretically, the effect of such isotropic shell deformations gives rise to local many-body correlations with neighboring particles, which in turn facilitates nearby particle collapses to ultimately yield a complete and discontinuous isostructural phase transition. 

It is our hope that the  coherent framework connecting the molecular architecture of the soft particles with their interfacial particle morphology and the resulting collective phase behavior not only provides a fundamental understanding of soft materials at interface, but also contributes to the discovery of new self-assembled structures by rationally targeting molecular architectures required for more complex, theoretically predicted assembly phases.

\section*{Materials and Methods}
\subsection*{Experiments}
\label{sec:experiments}
A detailed description of the PNIPAM-based core-shell particle synthesis by precipitation polymerisation can be found in previous work~\cite{ReySM17,Tang2018} and in the supplementary information. In short, PNIPAM microgel particles (continuous regime) with $1\,\mathrm{mol}\,\% $ crosslinker ($N$,$N'$-methylenebisacrylamide) (BIS) were synthesized in a one-pot reaction~\cite{ReySM17}. Silica-PNIPAM core-shell particles undergoing a heterostructural phase transition were obtained by surface functionalisation of Stöber silica cores (d$_{C}$=160 nm) with methacrylate moities to anchor the polymer shell to the surface. Then, a PNIPAM microgel shell ($5\,\mathrm{mol}\,\%$ BIS) was added onto the cores in a one pot, batch method.
For the core-shell particles undergoing an isostructural phase transition, we used the same synthesis as above, but continued to add additional monomer and crosslinker via a continous feeding process ($5\,\mathrm{mol}\,\%$ BIS)\cite{Tang2018}.
The interfacial phase diagram of the core-shell particles at the air/water interface was obtained by the simultaneous compression and deposition method~\cite{ReySM2016,ReySM17} using a Langmuir trough followed by characterisation using SEM (Zeiss Gemini 500). The morphology was investigated by AFM (JPK NanoWizzard) in AC mode using a MikroMash NSC-18/AL BS cantilever.

\subsection*{Simulation model: augmented ensemble} 
\begin{figure}
\centering
\includegraphics[width=0.5\textwidth]{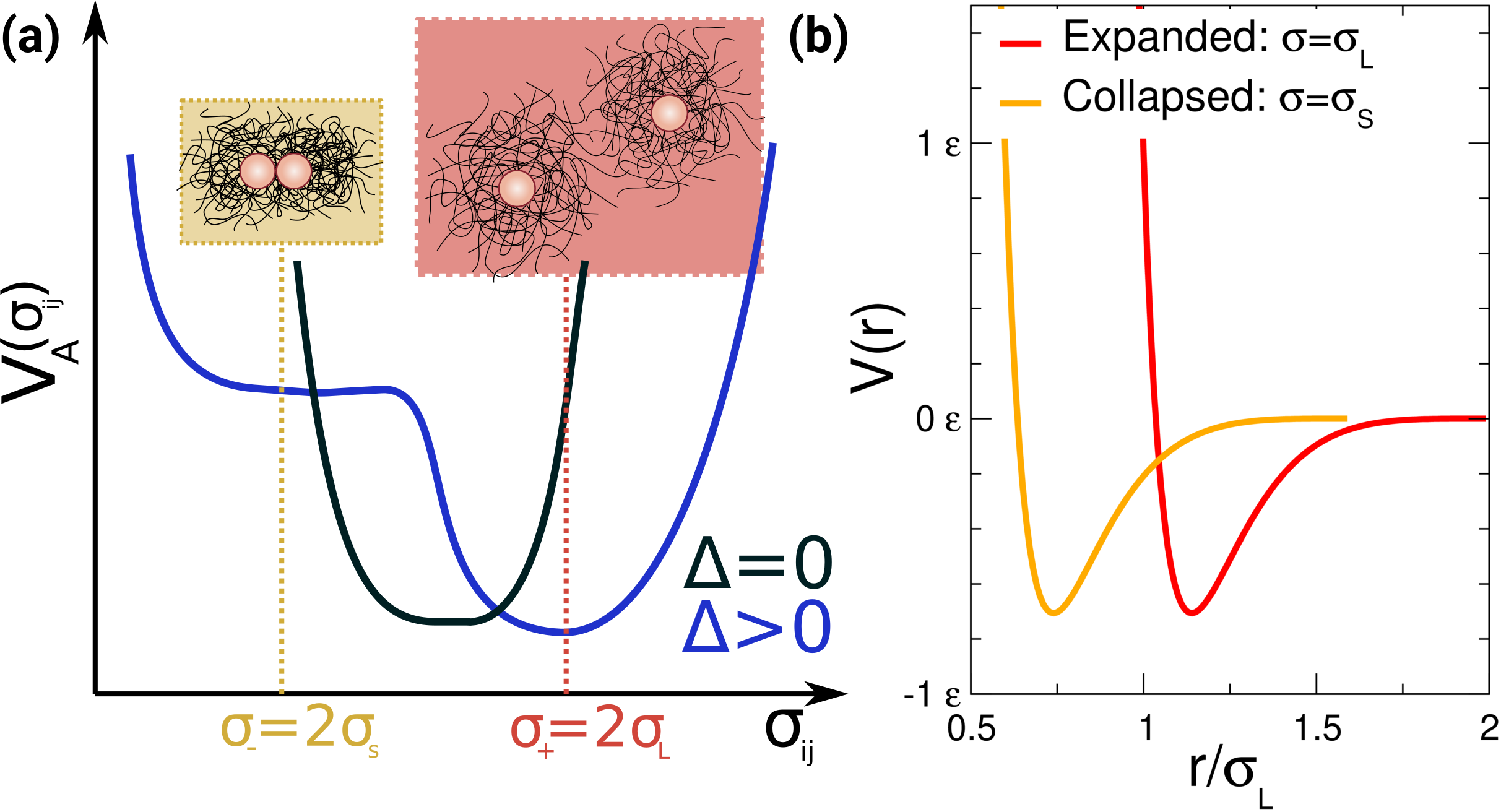}
\caption{(a) Augmented potentials as a function of the interaction distance $\sigma_{ij}$. 
The black line represents eq.~\ref{eq:dw} when $\Delta=0$, such that the potential is effectively $\left(\sigma_{ij}-\sigma_0\right)^{2\cdot4}$. 
The blue line corresponds to $\Delta>0$, provoking the appearance of a second length scale. 
The minimum at $\sigma_{+}=2\sigma_L$ represents the favorable situation of expanded particles while the metastable minimum at $\sigma_{-}=2\sigma_S$ is the collapsed state. (b) Pairwise potential defined in eq.~\ref{eq:lj} as a function of the central distance $r$. Since the interaction distance is treated as a variable, the potential can shift over time. See the Methods section for more details.}
\label{fig:pot}
\end{figure}

In our simulations, two particles separated by a distance $r_{ij}$ interact through the following potential~\cite{PhysRevE.99.012106}:
\begin{align}
\begin{split}
&V(r_{ij},\sigma_{ij})= \\ &= \left\{
\begin{array}{ll}
6\epsilon\left[\left(\frac{\sigma_{ij}}{r_{ij}}\right)^{12}-\left(\frac{\sigma_{ij}}{r_{ij}}\right)^{6} +\sum\limits_{l=0}^3 c_{2l}\left( \frac{r_{ij}}{\sigma_{ij}} \right )^{2l}\right], &\mathrm{if}~\frac{r_{ij}}{\sigma_{ij}}<x_c\\
0,&\mathrm{if}~\frac{r_{ij}}{\sigma_{ij}}\ge x_c
\end{array}
\right.
\label{eq:lj}
\end{split}
\end{align}
which is a modified Lennard-Jones (LJ)  potential with energy well depth $\epsilon$.
The parameter $\sigma_{ij}$ is the sum of the radius $\sigma_i$ of particle $i$ and the radius $\sigma_j$ of particle $j$. The potential is plotted in Fig.~\ref{fig:pot}(b) for different values of $\sigma_{ij}$.
The coefficients $c_{2l}$ are determined by requiring that the first three derivatives of $V$ vanish at the adimensional cutoff distance $x_c=2$.
These corrections to LJ are fundamental in the augmented ensemble, because the cutoff distance might be coupled to augmented variables and can thus change over time, preventing the use of constant tail corrections~\cite{Frenkel2002}. It is important to use an attractive potential in order to model the capillarity effect at the interface~\cite{Ciach2017}.

\subsection*{Augmented potentials}
\begin{figure}
\centering
\includegraphics[width=0.8\columnwidth]{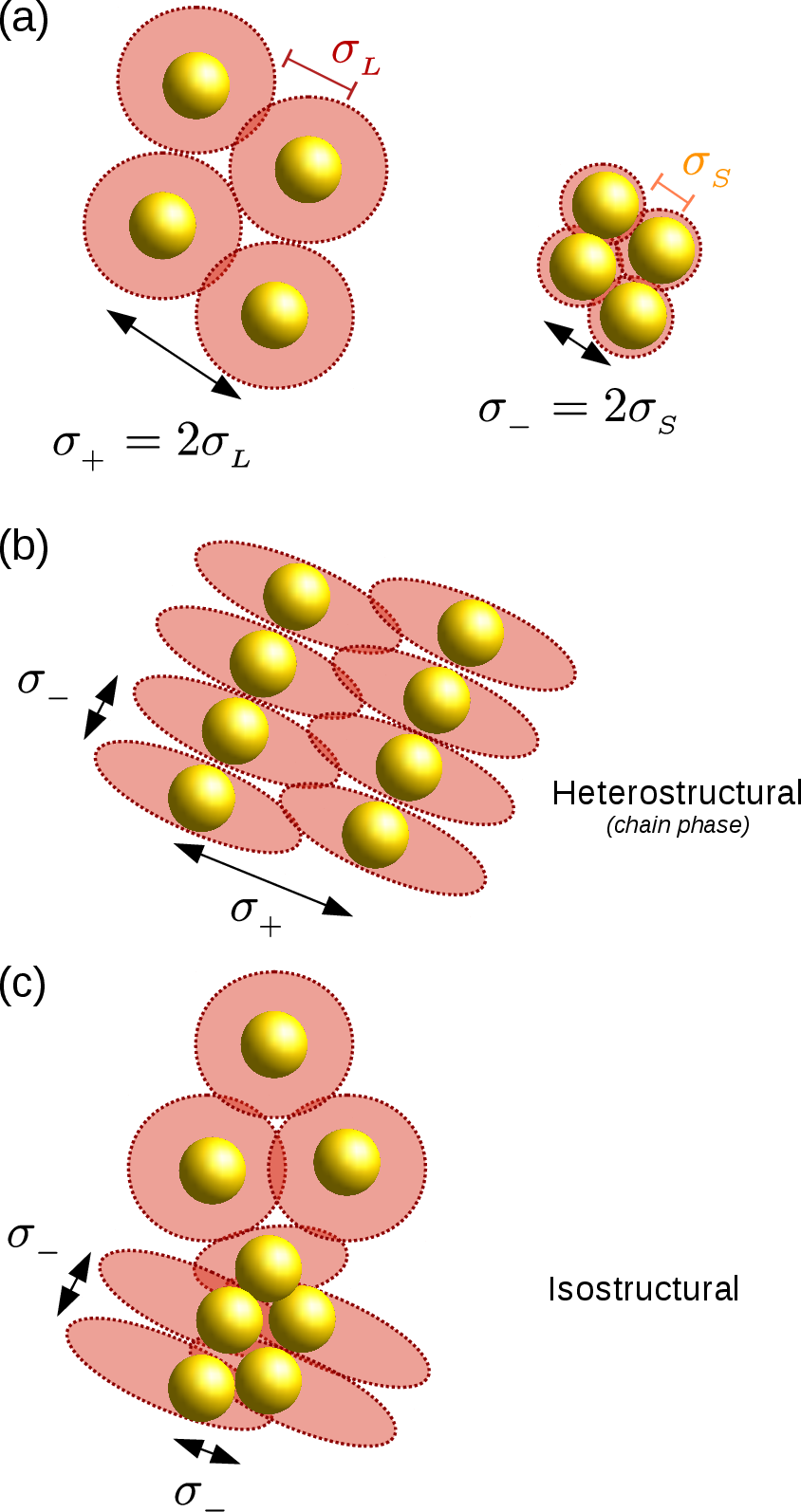}
\caption{Effect of the augmented potential on the particle size. (a) The stable state $\sigma_{ij}=\sigma_+$ corresponds to two expanded particles of size $\sigma_L$ in contact, while the metastable $\sigma_{ij}=\sigma_-$ corresponds to the contact of two collapsed particles of size $\sigma_S$. (b) In the heterostructural regime, chain-like structures are stable for intrachain interaction ranges of $\sigma_{ij}=\sigma_-$ and interchain interaction ranges of $\sigma_{ij}=\sigma_+$. (c) In the isostructural regime, settting $\lambda_m>0$, we impose that if a particle is collapsed in the chain direction it is also more likely to collapse with its other neighbors, creating the typical close-packing hexagonal domains.
}
\label{fig:rule}
\end{figure}
Augmented potentials and variables are phenomenologically built in order to closely reproduce experimental results, similarly to fictitious pairwise potentials~\cite{wca1971,Dijkstra1999}. In this section we explain the physics behind them. The first step is the choice of what augmented variables represent.
In the Supplementary Material we use the $\sigma_i$ as augmented variables, following Ref.~\cite{PhysRevE.99.012106,Berthier_2019}. We show that this approach is well suited to model particles that can expand and collapse, which is the case for, e.g., polymer stars, coils and brushes~\cite{Likos2008,Louis2002} and responsive particles~\cite{Geisel2012,Senff1999,PhysRevE.102.042602,baul2020structure}. 
However, the soft particles that we want to model here can collapse only when the reciprocal pressure of a pair of particles exceeds the threshold, so it is not possible to have a single collapsed particle between only expanded neighbors.
We model this feature by using the interaction distances $\sigma_{ij}$ as our augmented variables. If $\sigma_{ij}$ is large then we have a pair of widely spaced expanded particles, while if $\sigma_{ij}$ decreases the two particles can get closer as if they are collapsing.
Noticeably the dynamics of $\sigma_{ij}$ is coupled to the positions by the pairwise potential since eq.~\ref{eq:lj} does depend on  $\sigma_{ij}$. This produces a force acting on $\sigma_{ij}$ equal to $-\partial V(r_{ij},\sigma_{ij})/\partial\sigma_{ij}$. To bound the augmented variables in a physical regime, we add an external augmented potential $V_A(\sigma_{ij})$ that acts only on the augmented variables. The potential has to be phenomenologically built in order to reproduce the correct physics of the system. In particular we propose:
\begin{equation}
    V_A (\sigma_{ij})=K \left[ \left(\sigma_{ij}-\sigma_0\right)^4 -\Delta\right]^2-5K\Delta\left(\sigma_{ij}-\sigma_0 \right)~,
    \label{eq:dw}
\end{equation}
which is based on a standard double well potential with a barrier height of $K\Delta^2$ (represented by the first term of eq.~\ref{eq:dw}). The values of $\sigma_0$ and $\Delta$ control the position of two (or one for $\Delta=0$) minima $\sigma_{\pm}=\sigma_0 \pm \Delta^{1/4}$. 
Setting $\Delta=0$ is an easy way to model (in the augmented ensemble) particles whose stiffness increases with compression, the fundamental requirement of the continuous regime.
Alternatively, $\Delta>0$ introduces the two length scales required for the heterostructural and isostructural regimes.
For the simulations reported in Figures~\ref{fig:reg1},~\ref{fig:reg2} and~\ref{fig:reg3} we calculate the precise values of $\sigma_0$ and $\Delta$ that match the experimental measurements (measured and reported in the Sup. Info): $\sigma_0=1500\mathrm{nm}$ is the preferred size in the continuous regime, $\sigma_+=480\mathrm{nm}$ and $\sigma_-=170\mathrm{nm}$ are the two length scales in the heterostructural regime, and $\sigma_+=950\mathrm{nm}$ and $\sigma_-=500\mathrm{nm}$ for the isostructural case.
Consequently, we are able to see the effect of the compression at the same area per particle that we measure in the experiments.
Furthermore, we add to the double well potential a linear term in order to favor the configurations with $\sigma_{ij}\sim\sigma_{+}$ to model the fact that a pair of microgels prefers to stay at its expanded size, where the polymer chains have more space to move. 
The particular coefficient $-5K\Delta$ we use in eq.~\ref{eq:dw} is calculated imposing that $V_A(\sigma_-)-V_A(\sigma_+)=10K\Delta^2$ corresponding to $10-$times the energy barrier, such that the configurations with $\sigma_{ij}\sim\sigma_{-}$ are very short lived and thus only favorable at high compression. Lastly the prefactor $K$ represents the overall magnitude of the potential that dictates the energy scale and thus the stiffness of the particle.
Then, the augmented variables evolve following the augmented force $F_A$:
\begin{equation}
  - F_A(\sigma_{ij})= \frac{\partial V(r_{ij},\sigma_{ij})}{\partial\sigma_{ij}} + \frac{\partial  V_A(\sigma_{ij})} {\partial\sigma_{ij}}
\end{equation}
leading to the equation of motion
\begin{equation}
  m_{\sigma}\ddot{\sigma}_{ij}(t)= F_A(\sigma_{ij})~.
\end{equation}
where $m_{\sigma}$ is a fictitious mass for the $\sigma_{ij}$ dynamics.
In the Supplementary Material we write down the full (Pseudo)-Hamiltonian which governs
the equations of motion.

We plot the augmented potentials in Fig.~\ref{fig:pot}(a). It is clear that the energetically favored states correspond to $\sigma_{ij}\sim\sigma_{+}$. That value represents two particles in contact while being at their maximum size in the expanded state of radius $\sigma_L$, so $\sigma_{+}=2\sigma_L$. To collapse to the small size of $\sigma_{-}=2\sigma_S$ the pair has to spend some energy and cross the barrier. 
Numerically, $\sigma_+$ and $\sigma_-$ correspond to the average diameter of particles in the collapsed and expanded AFM pictures, respectively.
Furthermore  the $V_A$ collapsed state is made metastable by the linear term, such that particles always expand if not compressed.
Since $\sigma_{ij}$ can change, it means that the pairwise interaction in eq.~\ref{eq:lj} has to readjust according to the global and local environment probed by $\sigma_{ij}$. In this sense we show in Fig.~\ref{fig:pot}(b) that the equilibrium length becomes smaller for collapsed pairs.


Lastly, with the support of Fig.~\ref{fig:rule} we discuss the effect of many-body interactions to distinguish between the heterostructural and the isostructural simulations. 
Since every particle $i$ is involved in $N$ interactions with all the others $j$, the simplest way to define the size of each particle is through
\begin{equation}
    r_{i,\textrm{min}} = \min_j \frac{\sigma_{ij}}{2}.
\end{equation} 
We say that a particle is collapsed if $r_{i,\textrm{min}} \sim \sigma_S$, so $i$ has to reduce the value of $\sigma_{ij}$ with just one of its neighbors $j$ in order to be considered collapsed. As a consequence, even if $r_{i,\textrm{min}} \sim r_{j,\textrm{min}} \sim \sigma_S$ it is possible that $\sigma_{ij}\gg \sigma_-=2\sigma_S$. This means that two particles that are collapsed, might still have a long range repulsion that requires additional energy to eventually collapse.
From a microscopic point of view, this extra repulsion is attributed to the loosely crosslinked shell, which being characterized by long dangling polymer chains, is able to rearrange in order to occupy more space around the core-core contacts.
This effect (sketched in Fig.~\ref{fig:rule}(b)) leads to chain formation because particles want to assume a phase where the contacts are minimal, so all the energy is invested in collapsing only the 2 contacts per particle within the chains, producing the heterostructural regime.
Finally, to describe the isostructural regime we have to prevent this anisotropic effect. We do so by imposing that:
if $r_{i,\textrm{min}} + r_{j,\textrm{min}} < \lambda_m  \sigma_{-}  \longrightarrow \sigma_{ij}=\sigma_-$. We set $\lambda_m=0$ (no multi-body effects) to reproduce the results in Fig.~\ref{fig:reg1} and Fig.~\ref{fig:reg2}, while we set $\lambda_m=1.01$ to produce Fig.~\ref{fig:reg3}.
This additional $\lambda_m$-dependent rule mimics the effect of isotropic compression, such that a partial collapse induced by one neighboring particle will be felt by all neighbors; these in turn can then stimulate a full collapse. This also suppresses the chain phase that requires $\sigma_{ij}=\sigma_+$ to stabilize the interactions between neighboring chains, as we show in Fig.~\ref{fig:rule}(c). 

\subsection*{Computational details}
We perform two-dimensional simulations with periodic boundary conditions to model the interfacial layer.
All of our simulations are MD runs in the canonical $NVT$ ensemble, at constant particle number $N=500$ and temperature $T$; the volume $V=L^2$ is used as the control parameter. We use the leap-frog integration method coupled to a Berendsen thermostat~\cite{Berendsen1984} to keep $T$ constant, and we vary $L$ to approach the desired area per particle $A_P=L^2/N$. After equilibrating the interface for $10^6$ time steps of size $dt=10^{-3}$ reduced units at $T=2\epsilon$ and large $A_P$, we gradually compress the system reducing $L$ and in turn $A_P$.
We use the FIRE algorithm~\cite{FIRE} to minimize the energy and find the inherent structure at target $A_P$, that resembles the experimental quench that happens at the interface.
As discussed in the previous section, the value of $\Delta$ and $\sigma_0$ are dictated by the experiments: (i) $\sigma_0=1500\mathrm{nm}$ is the preferred size in the continuous regime (and $\Delta=0$ to have a single minimum), (ii) $\sigma_+=480\mathrm{nm}$ and $\sigma_-=170\mathrm{nm}$ set $\sigma_0$ and $\Delta$ for the the heterostructural regime, and lastly (iii) $\sigma_+=950\mathrm{nm}$ and $\sigma_-=500\mathrm{nm}$ determine $\sigma_0$ and $\Delta$ in the isostructural regime.
The stiffness parameter instead dictates the response of the microgels to compression and we tailor it to reproduce the experimental measurements by using the following values: $K=0.01\epsilon \mathrm{nm}^{-8}$ for the continuous regime (Fig.~\ref{fig:reg1}), $K=1\epsilon \mathrm{nm}^{-8}$ for the heterostructural regime (Fig.~\ref{fig:reg2}) and $K=0.05\epsilon \mathrm{nm}^{-8}$ for the isostructural regime (Fig.~\ref{fig:reg3}), where $\epsilon$ is the unit of energy. 
For the many-body interactions we set $\lambda_m=0$ to disable such effects in the continuous and heterostructural regimes, while we set $\lambda_m=1.01$ to obtain the isostructural regime.
The last unit that we set is the augmented mass $m_{\sigma}=m=1[M]$ that represents the inertia of the augmented variable, dictating how rapidly a pair can change $\sigma_{ij}$, and we sat that equal to the ordinary mass $m$. The importance of the augmented potential compared to the standard LJ interaction can be tweaked either through $K$ or $m_{\sigma}$, but we suggest to modify $K$ and let $m_{\sigma}=m$ to avoid numerical instabilities.
We repeat the procedure for $M> 50$ randomly generated configurations in order to verify the consistency of the results and get a better statistics for the histograms.

\section{Acknowledgements}

We thank J.S.J. Tang for her contributions to the particle synthesis and J. Horbach and J. Wang for helpful discussions. This work was funded by the Deutsche Forschungsgemeinschaft (DFG) under grant number VO 1824/8-1 and LO 418/22-1. N.V. also acknowledges support by the Interdisciplinary Center for Functional Particle Systems (FPS).

\section{Additional information}
\textbf{Supplementary information} is available for this paper at link.

\clearpage

\section*{Supplementary information}
\begin{figure}
\includegraphics[width=1\columnwidth]{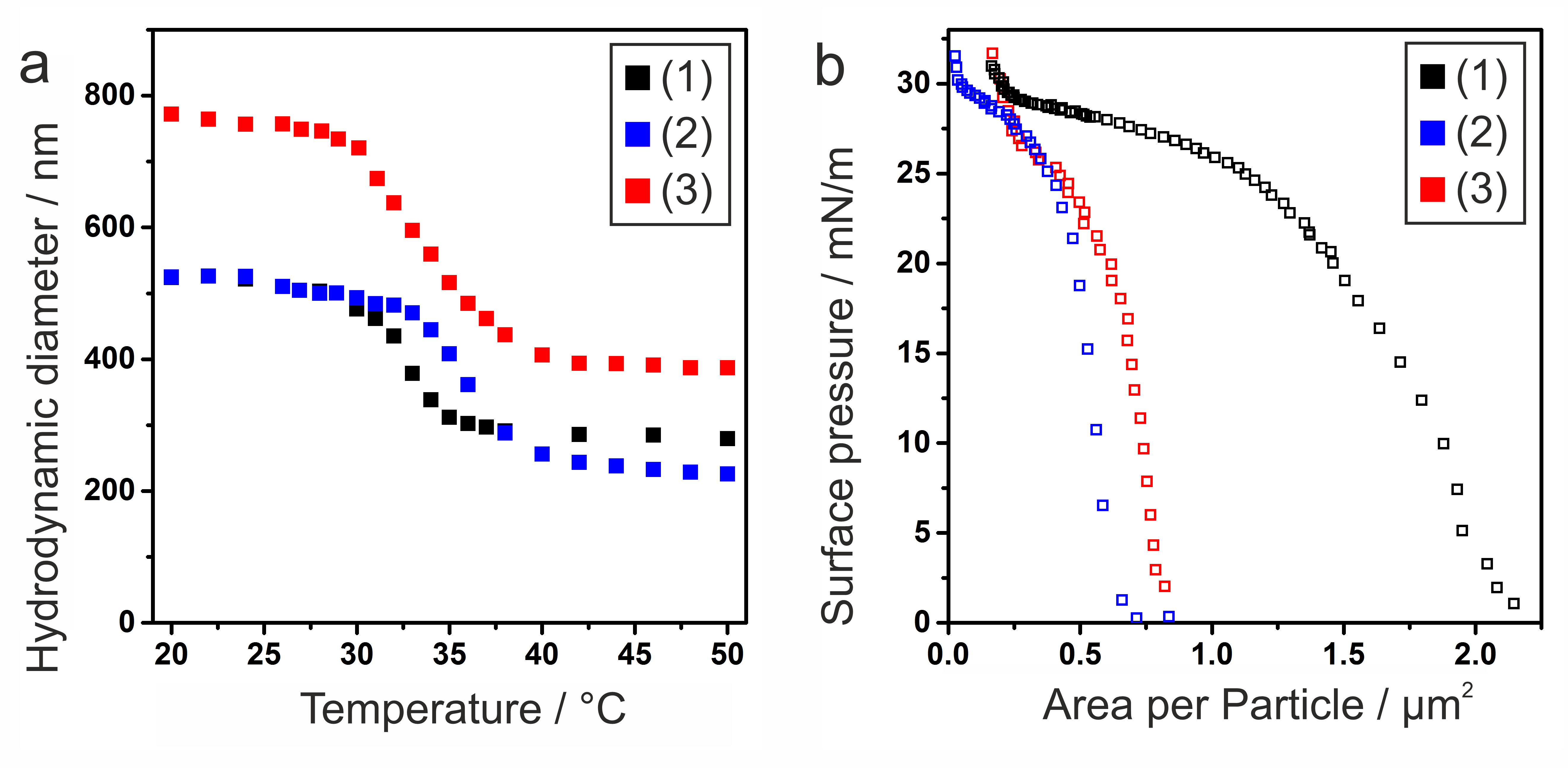}
\caption{Bulk and interfacial properties of the three core-shell particles. a) Hydrodynamic diameter - temperature diagram measured by dynamic light scattering and (b) surface pressure - area per particle diagram for core-shell particles undergoing a (1) \textit{continuous} (black), a (2)  \textit{heterostructural} (red) or a (3)  \textit{isostructural} (blue) phase transition.}
\label{fig:s3}
\end{figure}

\begin{figure*}
\includegraphics[width=0.8\textwidth]{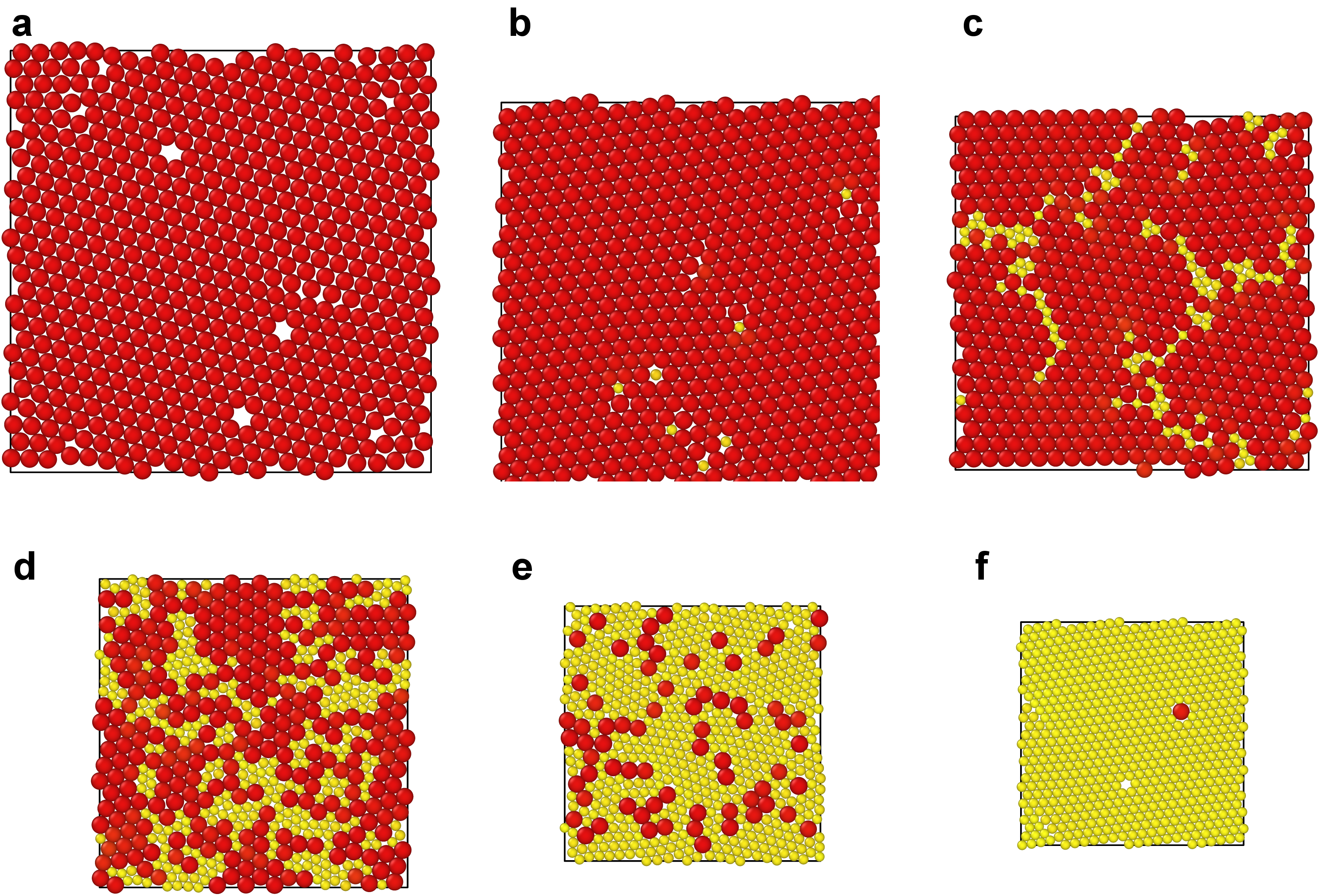}
\caption{Inherent structures of simulations in the augmented ensemble. The compression increases from (a) to (f). Simulated particles are red in their expanded size and they follow a gradient toward yellow while collapsing. They form a low packing hexagonal lattice when the area per particle is large (a), and they transition towards hexagonal close-packing upon compressing (f). This model can be framed in the \textit{isostructural} regime, despite not capturing the details of microgels or core-shell particles collapsing at an interface.
The use of sizes as augmented variables provokes the collapse of single particles rather than pairs, as proven by the presence of lonely collapsed (yellow) or expanded (red) particles from (b) to (f).}
\label{fig:s1}
\end{figure*}
\section{Size as augmented variable}
In this section we discuss a simpler implementation of the augmented ensemble, which could be useful to model many colloidal systems. For the sake of simplicity we consider here a model where only the particle sizes $\sigma_i$ are the augmented variable. The case considered in the main text where all pair interaction distances $\sigma_{ij}$ then follows from our considerations in a straight-forward way. 
It has been shown that this particular augmented ensemble can closely reproduce polydispersity~\cite{Brito2018,PhysRevE.99.012106,baul2020structure} 
and provide a significant speed up in the context of low temperature equilibration~\cite{Berthier_2019}. 
We argue here, that this augmented ensemble approach provides not only a way to model small variations in the particles shape that result in the inevitable experimental polydispersity, but it is also a profitable strategy in modelling complex structures. 

We propose an augmented potential acting on the particle size $\sigma_i$ with the following shape:
\begin{equation}
    V (\sigma_{i})=K \left\{b \left[ \left(\sigma_{i}-\sigma_0\right)^2 -\Delta\right]^2-\sqrt{2}\Delta^3\left(\sigma_{i}-\sigma_0 \right) \right\}
\end{equation}
based on a standard double well potential with a barrier height of $\Delta$, similarly to the augmented potential proposed in the main text.
The value of $\sigma_0$ controls the position of the two minima $\sigma_{\pm}=\sigma_0 \pm \Delta$, while the linear term coefficient $\sqrt{2}\Delta^3$ is chosen to impose $V'(\sigma_{-})=0$. Lastly the prefactor $K$ represents the overall magnitude of the potential and thus the stiffness of the particle.
So due to the unbalance created by the linear term $\sigma_{+}$ is stable while $\sigma_{-}$ is metastable.
This means that the configuration that minimizes the augmented energy corresponds to expanded particles with $\sigma=\sigma_{+}$, but for a large surface pressure $\Pi$ a particle can conveniently assume the state $\sigma=\sigma_{-}$ opening some space by reducing its size.

Then, the size of each particle evolves following the augmented force $F_A$:
\begin{equation}
  - F_A(\sigma_{i})= \sum_{i<j} \frac{\partial V(r_{ij},\sigma_{i},\sigma_{j})}{\partial\sigma_{i}} + \frac{\partial  V_A(\sigma_{i})} {\partial\sigma_{i}}
\end{equation}
where the particle-particle interaction $V(r_{ij},\sigma_{i},\sigma_{j})$ depends on both $\sigma_i$ the size of particle $i$ and $\sigma_j$ the size of particle $j$ at that timestep.
Each size evolves according to the equation of motion
\begin{equation}
  m_{\sigma}\ddot{\sigma}_{i}(t)= F_A(\sigma_{i})~.
\end{equation}

Now we derive the (pseudo)-Hamiltonian in the case that augmented variables $\xi_i$ are added to the positions $r_i$ and momenta $p_i$.
We use $\xi_i$ to enclose in the same notation both the sizes as augmented variables ($\sigma_i$) and the pairwise interactions $\sigma_{ij}$ that needs to be used to realistically model two microgel particles that collapse, as we explain in the main manuscript.
Then the (pseudo)-Hamiltonian which governs the equation of motion is
\begin{equation}
	H(p_i,r_i,\xi_{i},p_{\xi_{i}})= H_{\mathrm{kin},1}(p_i) + H_{\mathrm{kin},2}(p_{\xi_{i}})+V_\mathrm{pot}(r_i,\xi_{i})~,
\end{equation}
where
\begin{equation}
	H_{\mathrm{kin},1}(p_i)=\sum_{i=1}^N \frac{p_i^2}{2m}
\end{equation}
is the kinetic energy of the particles and
\begin{equation}
	H_{\mathrm{kin},2}(p_{\xi_{i}})=\sum_{i=1}^{N_{\xi}} \frac{p_{\xi_{i}}^2}{2m_{\xi}}
\end{equation}
is the (fake) kinetic energy of the augmented variables. If the mass $m_{\xi}$ is sufficiently small, the sizes will practically follow the the microgel coordinates adiabatically similar in the spirit to what the electronic degrees of freedom follow the ion coordinates in a Car-Parrinello-like simulation~\cite{car1985}.
The full potential energy is
\begin{equation}
    V_\mathrm{pot}(r_i,\xi_{i})=\sum_{i,j=1}^N \left( V(r_{ij},\xi_i,\xi_j) + V_A(\xi_i,\xi_j)\right)~.
\end{equation}
Notice that using augmented variables that depend on two particles like we do in the main manuscript ($\xi=\sigma_{ij}$), leads to $N_\xi=N^2$ and thus a significant increase in the number of degrees of freedom. However since the interactions are only short-ranged this increase in the computational demand can be mitigated by techniques like neighbour, cell or Verlet lists~\cite{Frenkel2002}. Overall, this suggests that one should be careful in defining the correct augmented variables that are better suited to model the desired phenomenon.  

It is important to note here that though all the original interaction between the microgels are pairwise, the presence of the augmented variables lead to effective many-body potentials between the microgel centers if the augmented variables are integrated out. Since a Hamiltonian of the full augmented system exists, statistics can be applied in the standard way for the full system and then reduced to the microgel centers by using the classic concept of effective interactions.  

Lastly, to produce the results reported in Fig.S~\ref{fig:s1}, we follow a protocol similar to the one explained in the Methods section of the main paper, but here we model the compression by changing the surface pressure $\Pi$ through a Berendsen barostat. At low surface pressure (a), all the particles assume the $\sigma=\sigma_{+}$ state and for this reason we color them in red. Increasing $\Pi$ (b), it is more convenient to collapse only some particle to $\sigma=\sigma_{+}$ (colored in yellow) rather than compressing a bit all of them. This is true because we choose a relatively high $K$ that effectively models stiff particle that do not like to be compressed but can collapse to a smaller size if compressed beyond a threshold. This implementation of the collapse is naturally reversible.
Many colloidal systems like polymer stars, coils and brushes~\cite{Likos2008,Louis2002} and responsive microgels~\cite{Geisel2012,Senff1999} have a similar mechanism.
Compressing the system even more (c)-(e), clusters of collapsed particles grow, until all the particle in the system are collapsed forming a close-packed hexagonal lattice (f). 
Following the classification introduced in the Main paper, we can classify this system within the \textit{isostructural} regime.

This simpler augmented scheme does not represent microgels and core-shell particles at an interface because of the nature of a collapse event. In fact, the structure of each particle is such that a collapse event always involve two particles. When those two particles apply a reciprocal pressure beyond a threshold, they rapidly collapse putting their cores in contact. So in the systems that we discuss in the Main paper, it is not possible to have a single collapsed particle as in Fig.S~\ref{fig:s1}(b)-(f).

\section{Experimental details}

\textbf{Materials:} All chemicals were obtained from commercial sources and used as received if not stated otherwise. N,N’-Methylenebis(acrylamide) (BIS; 99 $\%$, Sigma Aldrich), ethanol (EtOH, 99.9 $\%$, Sigma Aldrich), ammonium persulfate (APS; 98 $\%$ Sigma Aldrich), tetraethyl orthosilicate (TEOS; 98 $\%$, Sigma Aldrich), ammonium hydroxide solution (28-30 $\%$ NH$_{3}$ basis, Sigma Aldrich), (3-(trimethoxysilyl)propyl methacrylate (MPS; 98 $\%$, Sigma Aldrich) were used as received. N-Isopropylacrylamide (NIPAM; 97 $\%$, Sigma Aldrich) was purified by recrystallization from hexane (95 $\%$, Sigma Aldrich). Water was double deionized using a Milli-Q system (18.2 $M\Omega$.cm, Elga\texttrademark~ PURELAB\texttrademark~ Flex). 

\textbf{Synthesis:} 
Colloidal silica particles used as cores with a diameter of $160$ nm $\pm 10$ nm were prepared according a modified St\"ober process\cite{Tang2018}. In a round bottom flask, 250 mL EtOH , 12.5 mL Milli-Q water and 25 mL NH$_{3}$ (aq) were stirred together. 18.75 mL of TEOS was stirred in 75 mL EtOH and both solutions were heated to $50$ \textdegree C and equilibrated for 30 min. The TEOS solution was then quickly added to the first mixture under heavy stirring. As soon as the reaction mixture became turbid, the prepared fluorescent dye solution was slowly added. We let the reaction proceed for 2 d at $50$ \textdegree C. 
This suspension was used for functionalization without any further purification by adding 102.7 $\mu L$ MPS. We allowed the reaction mixture to stir at room temperature for at least 1 d and then boiled it for 1 h to ensure successful functionalization. Afterwards, we purified the particles by centrifugation and redispersed them three times in ethanol and three times in Milli-Q water. 

(1) The PNIPAM microgel undergoing a continuous phase transition were synthesised by surfactant-free precipitation polymerization according to previous work\cite{ReySM17}. In a 500 mL three-neck round bottom flask equipped with reflux condenser and stirrer, 2.83 g of NIPAM and 38.6 mg BIS ($1\mpc$) were dissolved in 249 mL of Milli-Q water. The solution was heated to $80$ \textdegree C and purged with nitrogen gas. After an equilibration time of 30 min, the nitrogen gas inlet was replaced by a nitrogen-filled balloon to sustain the nitrogen atmosphere. The reaction was initiated by rapidly adding a solution of 14.3 mg of APS dissolved in 1 mL of Milli-Q. After 5 h the turbid mixture was slowly cooled down to room temperature. The suspension was purified by centrifugation and redispersion in Milli-Q water three times, followed by dialysis against Milli-Q water for one week. 

(2) SiO$_{2}$-PNIPAM core-shell particles undergoing a heterostructural phase transition were obtained by growing a PNIPAM shell onto the silica cores via a batch surfactant-free precipitation polymerization. In a 500~mL three-neck round bottom flask, 282.9 mg NIPAM and 19.3 mg BIS ($5\mpc$) were dissolved in 47 mL Milli-Q water. We added the 2.591 g aqueous SiO$_2$ core dispersion (6.6 wt$\%$). The solution was heated to $80$ \textdegree C and purged with nitrogen. After equilibration for 30 min, a balloon filled with nitrogen was used to keep the nitrogen atmosphere. Subsequently, 11 mg APS was rapidly added to initiate the reaction. We let the reaction proceed for 4 h and after it cooled down, we purified the suspension 6 times by centrifugation and redispersion in Milli-Q water. 

(3) SiO$_{2}$-PNIPAM core-shell particles undergoing an isostructural phase transition were obtained by growing a PNIPAM shell onto the silica cores via a semi-batch surfactant-free precipitation polymerization adapted from our previous report\cite{Tang2018}. In a round bottom flask, 282.9 mg of NIPAM and 19.3 mg BIS were dissolved in 45 mL Milli-Q water. We added the 3.767~g aqueous SiO$_{2}$ core dispersion (6.6 wt$\%$). The mixture was heated to $80$ \textdegree C and purged with nitrogen. After an equilibration time of 30 min, the nitrogen gas inlet was replaced by a nitrogen-filled balloon to sustain the nitrogen atmosphere. The reaction was started by rapidly adding 114 mg APS dissolved in 1 mL Milli-Q water. We let the reaction proceed for an hour, before adding 114 mg of APS dissolved in 1 mL Milli-Q water again. Subsequently, a nitrogen purged solution of 2.037 g NIPAM and 138 mg BIS was added using a syringe pump (3 mL/h) for further shell growth. 2 h after the last addition of reagents, the reaction mixture was allowed to cool to room temperature. The core-shell particles were purified by 10 times centrifugation and redispersion with Milli-Q water to dispose of free microgel particles. 

Characterization: Dynamic light scattering was performed with a Malvern Zetasizer Nano-ZS in disposable polystyrene cuvettes at a scattering angle of $173$ \textdegree. The hydrodynamic diameter of the particles was measured as a function of the temperature, ranging from 20 to 50 °C. At each temperature the measurements were performed four times after an equilibration of 15 min.

Simultaneous compression and deposition: We used the simultaneous compression and deposition method to transfer the interfacial arrangement onto a solid substrate as described in previous work\cite{ReySM2016,ReySM17,Rauh2017,Tang2018}. We used a Teflon® Langmuir-Blodgett trough (KSVNIMA) (area = 550 cm$^{2}$, width = 7.5 cm) with Delrin® barriers and the surface pressure was measured by a Wilhelmy plate. Silicon wafers (Siltronix®) were cut to 8 x 1 cm$^{2}$ and cleaned by ultrasonication in ethanol and Milli-Q water, followed by oxygen plasma (Diener). The substrate was mounted in a 45° angle and the trough was filled with Milli-Q water. The core-shell particle suspension was diluted to 0.1 wt$\%$ and mixed with 30 wt$\%$ ethanol as spreading agent. The core-shell particles were spread on the air/water interface of the Langmuir-Blodgett trough using a regular 100 µL pipette. After 10 min of equilibration, the barriers were compressed with a speed of 4 mm/min and the substrate was lifted with a speed of 0.4 mm/min. The core-shell particle assembly was characterized by scanning electron microscope (SEM, Zeiss Gemini 500) using the inlens detector, a 30 µm aperture and a voltage of 1 kV.

AFM phase and height analysis: The height and phase images of the deposited core-shell particles were extracted by atomic force microscopy (AFM) using a JPK NanoWizard instrument in AC mode with a MikroMasch NSC-18/AL BS cantilever (resonance frequency 75 kHz, spring constant 2.8 N/m). We scanned regions of 2 x 2 µm$^{2}$ with a resolution of 512 x 512 pixels$^{2}$. The images were post processed with Gwyddion, using flattening, scar correction and median of differences fits.

\end{document}